# Larger holes as narrower degree distributions in complex networks

Kiri Kawato[a], Yukio Hayashi[a]

[a]*Japan Advanced Institute of Science and Technology,
1-1 Asahidai, Nomi, Ishikawa, 923-1292, Japan*

**Abstract**

Although the analysis of loops is not so much because of the complications, it has already been found that heuristically enhancing loops decreases the variance of degree distributions for improving the robustness of connectivity. While many real scale-free networks are known to contain shorter loops such as triangles, it remains to investigate the distributions of longer loops in more wide class of networks. We find a relation between narrower degree distributions and longer loops in investigating the lengths of the shortest loops in various networks with continuously changing degree distributions, including three typical types of realistic scale-free networks, classical Erdös-Rényi random graphs, and regular networks. In particular, we show that narrower degree distributions contain longer shortest loops, as a universal property in a wide class of random networks. We suggest that the robustness of connectivity is enhanced by constructing long loops of $O(\log N)$.

*Keywords:* Network science, Shortest loops, Continuously changing degree distributions, Topological connection structure, Ramanujan graphs

## 1. Introduction

In network science with the decades of groundbreaking from classical graph theory, there is not much research on loops. Here, we use the words "loop" in physics, while the word "cycle" is more common in computer science and graph theory. In particular, the loop whose inside is empty means a "hole". For example of the difficulty of studying loops, many theoretical approaches e.g. by using generating functions assume to be locally tree-like in analyzing the robustness of connectivity [1, 2, 3, 4, 5]. The tree-like assumption involves recursive calculations, therefore the explicit form of the solution is not obtained. In addition, the recursive calculations give approximative solution or fail to converge, when loops exist. Even in the case of lattices, to derive the exact solutions of percolation threshold requires a complex integral [6]. Moreover, complicated message-passing methods are applied to estimate approximative solutions with higher accuracies for percolation, graph spectra, and community detection in considering combinations of primitive directed cycles that do not contain any shorter directed cycles [7]. To overcome the difficulties for investigating loops, the equivalence of dismantling and decycling problems [8] is attractive in a wide class of random networks with light tailed degree distributions, although they are critical node detection [9] and the minimum feedback vertex set problems [10], respectively, known as NP-hard in computer science. Because the equivalence suggests that the optimally tolerant structure against the worst case of node removals tend not to be tree. In other words, it is crucial to enhance loops for improving the robustness of connectivity even in the intractability of exact optimal to construct the most tolerant network [8, 9, 10]. Thus, approximative methods may contribute to open the door for a new direction of research. In fact, by heuristic rewiring methods for enhancing loops, it has been commonly found that [11] the robustness becomes stronger as degree distributions are narrower. Moreover, a smaller variance of degree distributions improves the robustness of connectivity against both degree-targeted attacks and belief-propagation-based loop destruction attacks [12]. In adding links, longer loops are more dominant than shorter ones in enhancing the robustness against intentional attacks [13]. However, as a universal property, various networks with a strong modular structure become extremely vulnerable [14], even from the optimal structure of random regular networks without modular structure [15]. Moreover, such vulnerability also tends to occur in geographical networks with local modules (or communities) embedded on planer spaces [16]. The average length of the shortest paths tends to be not $O(\log N)$ in planer networks, which therefore does not have small-world property. In other words, the smaller variance particularly contributes to improving the robustness in the randomized networks without modular structures. The importance of loops is also discussed in proposing new measures of centrality based on loops [17, 18]. Based on the above motivations, we aim to find a relation between narrower degree distributions and longer loops to strengthen the robustness as follows.

It has been well-known that the shorter loops such as triangles are included in scale-free (SF) networks of social, technological, and biological systems [19, 20]. In particular, social networks have high clustering coefficients [5]. However, the numbers of the shortest quadrilateral, pentagonal, and larger polygonal loops are unclarified for complex networks, though it is possible to exactly calculate the number of very short loops (in whose inside nodes and links can exist) with the limited lengths from three to seven for a network [21, 22]. Although an approximate analytical framework for the cumulative distributions of the shortest loops has been developed [23] for networks with arbitrary degree distributions, the several numerical results cannot be directly compared because of different sizes $N$ as the total number of nodes or different the average degrees $\langle k \rangle = 2M/N$ ($M$ denotes the total number of links in a network) for only three types of realistic SF networks [19, 24], classical Erdös-Rényi (ER) random graphs [25], and regular networks. Even though the framework [23] is widely applicable to general random networks, the lengths of the shortest loops have not been related to the robustness in various networks with continuously changing degree distributions from power-law, exponential, nearly Poisson, to nearly unimodal, which include typical three types of SF networks, ER random graphs, and regular networks. As a universal property independent of detail differences in degree distributions, we find that the smaller variance lead to longer average length of the shortest loops. This finding suggest the significance that the robustness of connectivity is enhanced by constructing long loops of $O(\log N)$.

On the other hand, Ramanujan graphs as the special cases of $d$-regular graphs that they have a good property of highly tolerant to bisections and that the girth defined by the minimum length of loops is asymptotically $O(\log_{d-1} N)$ at $N \to \infty$ [26, 27]. Moreover, random regular networks are very close to Ramanujan graphs at large sizes [28], and also correspond to the optimally tolerant network against intentional attacks [29, 30], and cascading failures [31] by overloads. We show that, in random regular networks, the average length of the shortest loops is the same $O(\log_{d-1} N)$ as in Ramanujan graphs [26, 27]. This suggests that such loop length are not too large to maintaining the connectivity even against intentional attacks.

The organization of this paper is as follows. In Section 2, we describe how degree distributions are controlled by a parameter value of attachment probability. In Section 3, we show the distributions of the shortest loops in networks with various degree distributions. In Section 4, we summarize the results that the lengths of the shortest loops become longer as degree distributions are narrower, and discuss future works.

## 2. Method for investigating the shortest loops in complex networks

We investigate the lengths of the shortest loops in synthetic networks whose degree distributions are continuously changing. Subsection 2.1 explains how to generate these networks. Subsection 2.2 describes the calculation method for the lengths of the shortest loops. Here, the shortness of loop is measured by hops passing through connected nodes by links in a network.

*2.1. Networks with continuously changing degree distributions*

We consider networks with degree distributions $P(k)$ that are continuously changing between power-law, nearly Poisson, and nearly unimodal in well-known topological structure such as realistic scale-free (SF) networks [19, 24], classical Erdös-Rényi (ER) random graphs [25], and regular networks, respectively. There are several well-known growing SF network models, such as Barabási-Albert (BA) [19, 24], Price [32], and other models [33, 34], in which the power-law exponent $\gamma$ is tunable. However, we consider not only SF networks but also the wide class of randomized networks with the above continuously changing degree distributions. Although infinitely many degree distributions are mathematically possible, the only configuration model [35] for a given $P(k)$ cannot prohibit self-loops at a node and multiple links between two nodes in some cases. To easily obtain various degree distributions under such constraints, we first generate connected networks without self-loops and multiple links, and then randomize them to investigate the pure effect of $P(k)$ on the distributions of the shortest loops and their average length, as mentioned later. Thus, we use growing network (GN) [36] and inverse preferential attachment (IPA) [37] models for generating networks with continuously changing degree distributions. There is no other method for generating growing networks with the above mentioned degree distributions at least in the state-of-the-art.

Both GN and IPA models can be unified as follows. First, an initial network is set, e.g. as a complete graph of $m$ nodes. Here, $m$ is a constant integer number. Then, at each time step $t = 1, 2, 3, \ldots$, a new node is added and connected to $m$ existing nodes chosen by the attachment probability defined as proportional to $k_i^\nu$, where $k_i$ is the degree of node $i$, and $\nu$ is a parameter. This process is repeated until reaching a size $N$. Note that the total number of links is $M \approx mN$ except the initial number of links, and the average degree $\langle k \rangle = 2M/N \approx 2m$. The continuously changing degree distribution depends on the value of parameter $\nu$ in GN and IPA models as follows.

- $\nu > 0$: The degree distribution becomes wider by preferential attachment. At $\nu = 1$, it becomes a pure power-law in SF networks [24].
- $\nu = 0$: The degree distribution becomes exponential by random attachment [24].
- $\nu = -1$: The degree distribution becomes a nearly Poisson in ER random graphs.
- $\nu < 0$: The degree distribution becomes narrower by inverse preferential attachment.
- $\nu \to -\infty$: The generated networks approaches regular, however does not become exactly regular [37].

In IPA model, chain-like structures emerge when $\nu \ll 0$ [37]. Thus, to investigate the pure effect of degree distributions, we randomize the above generated networks by using the configuration model [35], which eliminates chain-like structures, degree-degree correlations, and other structures. First, after generating a networks by using GN and IPA models, each link is cut into two free-ends in a network. Then, there are $k_i$ free-ends of links emanated from a node $i$. A pair of free-ends is randomly chosen and connected. Since this process neither adds nor removes links at any node, the degree of each node is maintained for a given $P(k)$ in the generated networks by using GN and IPA models.

Figure 1 shows the degree distributions $P(k)$ in the generated networks for $N = 50000$ and the parameter $\nu = 1, 0.5, 0, -1, -5, -20$, and $-100$, whose cases are colored by gold, light blue, orange, purple, green, blue, and red lines, respectively. In particular, the case of $\nu = 1, 0, -1$, and $-100$ in Figure 1 correspond to power-law (SF networks), exponential, nearly Poisson (ER random graphs), and nearly unimodal (regular networks) degree distributions, respectively. We remark that the degree distributions $P(k)$ become narrower as $\nu$ decreases. Hereafter, we simply refer to ER random graphs and regular networks as the correspondence to nearly Poisson and nearly unimodal.

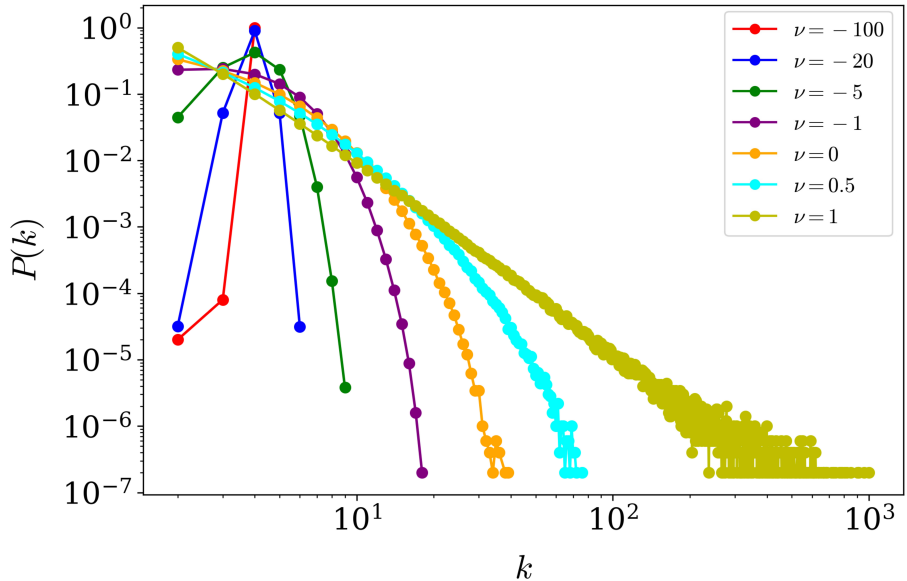

Figure 1: Degree distributions $P(k)$ in the generated networks for the number of $N = 50000$ nodes and $M \approx m \times N = 100000$ links ($m = 2$). Note that the average degree $\langle k \rangle \approx 2 \times m = 4$ is constant. Gold, light blue, orange, purple, green, blue, and red lines denote parameters from $\nu = 1$ to $\nu = -100$ in corresponding to SF networks, ER random graphs, and regular networks, respectively. The width of $P(k)$ becomes narrower as $\nu$ decreases from SF networks (gold line at $\nu = 1$) to regular networks (red line at $\nu = -100$).

*2.2. Calculation of the lengths of the shortest loops*

We describe the calculation method for the lengths of the shortest loops to which each link belongs. This is not the node-centric [23] but the edge-centric method. In randomized networks of finite size, since triangular loops of length three can exist at least with low frequency even when the size is large, these networks have the same girth = 3 therefore cannot be compared. Instead of the girth, with finite sizes, we aim to investigate the distributions of the shortest loops and their average length that may be related to the robustness of connectivity in various networks with continuously changing degree distributions [12]. As illustrated in Figure 2, the shortest loop consists of a link $e_{ij}$ and the shortest path between its end-nodes $i$ and $j$. The loop does not contain any loops in its inside because of the shortest. Thus, we calculate them as follows.

**Step 1.** Each link $e_{ij}$ between nodes $i$ and $j$ is temporarily removed from the network.
**Step 2.** The length $l$ of the shortest loops is obtained by adding 1 of link length of $e_{ij}$ to the length of the shortest path between nodes $i$ and $j$.
**Step 3.** Restore the removed link $e_{ij}$ to the network.
**Step 4.** Repeat Steps 1–3 for all links in the network.

By calculating the frequency of the length $l$ of the shortest loops, we obtain the distribution $P(l)$. The average length of the shortest loops is defined by $\langle l \rangle = \sum_l l P(l)$.

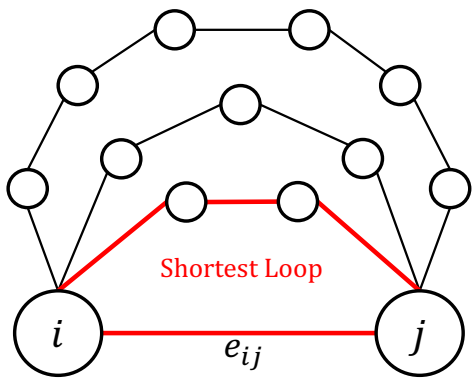

Figure 2: Illustration of the shortest loop (red lines) that consists of a link $e_{ij}$ and the shortest path between nodes $i$ and $j$. Note that the inside of loop is empty.

## 3. Results for the lengths of the shortest loops

We investigate the shortest loops in networks with continuously changing degree distributions $P(k)$, and show that the lengths of the shortest loops become longer in networks as $P(k)$ is narrower in the order from SF networks, ER random graphs, to regular networks. Remember that this order coincides with improving the robustness of connectivity against attacks [12]. In Subsection 3.1, we investigate the lengths of the shortest loops in real networks. In Subsection 3.2, we show that the variance $\sigma^2$ of $P(k)$ is a dominant factor on the average length of the shortest loops. The following results are averaged over 100 realizations for each combination of parameters $m = 2, \nu = 1, 0.5, 0, -1, -5, -20$, and $-100$.

*3.1. The lengths of the shortest loops in real networks*

In many real networks, SF structure commonly exists [19, 20] with power-law degree distributions which consists of a few high-degree hubs and many low-degree nodes. Moreover, SF networks are known to contain a large number of triangles [38, 39]. As a preliminary step before investigating longer loops, we investigate the number of triangles in these networks, which include not only SF networks but also ER random graphs, regular networks, and the intermediates between each pair of them.

We calculate the global clustering coefficient $C$ for a network. It is defined as follows [39]:

$$C_i = \frac{3 \times \text{(number of triangles)}}{\text{number of connected triples of nodes } i}, \quad (1)$$

$$C = \frac{1}{N} \sum_{i=1}^{N} C_i. \quad (2)$$

Here, the number of connected triples is defined by the sum of $\binom{k_i}{2}$ for node $i = 1, 2, \ldots, N$ with degree $k_i$. We denote $C_{\text{real}}$ and $C_{\text{expect}}$ calculated by Eqs. (1) and (2) for real networks and the randomized ones by using the configuration model [35], respectively. The datasets include several networks of AirTraffic [40], E-mail [41], Hamster [40], UCIrvine [40, 42], and Polblogs [43]. Note that only the largest connected components extracted from each of original datasets are analyzed. Examples of biological networks are omitted because randomization by the configuration model is difficult in such cases. Figure 3 shows the power-law exponent $\gamma$ estimated by the least-squares method.

Table 1(a) shows the global clustering coefficient $C$ for synthetic networks generated by using GN and IPA models with randomization for $N = 1500$, which is a similar size to real networks in Table 1(b). As the value of $\nu$ increases, $C$ becomes larger. In particular, SF networks ($\nu = 1$) with randomization contain more triangles than others for both ER random graphs ($\nu = -1$) and regular networks ($\nu = -100$). In other words, networks with the larger variance $\sigma^2 = \langle k^2 \rangle - \langle k \rangle^2$ of degree distributions $P(k)$ contain more triangles. Note that STD values slightly increases as larger $\nu$. We remark that the variance $\sigma^2$ is considered as the dominant factor on $C_{\text{expect}}$ for $\langle l \rangle_{\text{expect}}$ as shown later in Figure 5. In addition, $C_{\text{real}}$ is larger than $C_{\text{expect}}$. The reason may be caused from that the randomization eliminates such as degree-degree correlations and modular structure.

Figure 4 shows the distributions of the lengths $l$ of the shortest loops in real and the randomized networks. The distributions $P(l)$ and the cumulative distributions $P_{\text{SL}}(L>l)$ show that E-mail, Hamster, UCIrvine, and Polblogs contain shorter loops, except AirTraffic. Figure 5 shows that the average length $\langle l \rangle_{\text{expect}}$ of the shortest loops has a monotonically decreasing relation with the variance $\sigma^2$. These results show that larger values of $\sigma^2$ give larger $C_{\text{expect}}$ and smaller $\langle l \rangle_{\text{expect}}$. Note that the fitting curves in Figure 5 are estimated by the least-squares method for the function $a_3/\log(a_1 x^2 + a_2) + a_4$.

Table 1: The average values of the global clustering coefficient $C$ and its standard deviation (STD) calculated over 100 realizations of synthetic and real networks. (a) Synthetic networks are generated by using GN and IPA models with randomizations for $N=1500$, under $\langle k \rangle \approx 4$. As $\nu$ increases, the variance $\sigma^2$ of degree distributions $P(k)$ and $C$ becomes larger with a peak at $\nu=1$ as in SF networks. (b) Larger values of $\sigma^2$ give larger $C_{\text{expect}}$ and smaller $\langle l \rangle_{\text{expect}}$.

(a)

| $\nu$ | $\sigma^2$ | $C$ | STD |
|---|---|---|---|
| -100 | 0.01 | 0.0014 | 0.0007 |
| -20 | 0.10 | 0.0016 | 0.0006 |
| -5 | 0.88 | 0.0017 | 0.0008 |
| -1 | 3.24 | 0.0025 | 0.0009 |
| 0 | 5.90 | 0.0033 | 0.0009 |
| 0.5 | 9.57 | 0.0047 | 0.0008 |
| 1 | 30.02 | 0.0126 | 0.0018 |

(b)

| Network | | $N$ | $\langle k \rangle$ | $\sigma^2$ | $\gamma$ | $C_{\text{real}}$ | $\langle l \rangle_{\text{real}}$ | $C_{\text{expect}}$ | $\langle l \rangle_{\text{expect}}$ |
|---|---|---|---|---|---|---|---|---|---|
| Technological | AirTraffic | 1226 | 3.9 | 13.47 | 1.92 | 0.0639 | 4.30 | 0.0084 | 5.20 |
| Social | E-mail | 1133 | 9.6 | 87.23 | 1.46 | 0.1663 | 3.15 | 0.0274 | 3.65 |
| | Hamster | 1788 | 14.0 | 440.86 | 1.39 | 0.0904 | 3.20 | 0.0643 | 3.28 |
| | UCIrvine | 1893 | 14.6 | 599.57 | 1.33 | 0.0568 | 3.18 | 0.0829 | 3.14 |
| | Polblogs | 1222 | 27.4 | 1474.67 | 1.07 | 0.2260 | 3.01 | 0.1460 | 3.04 |

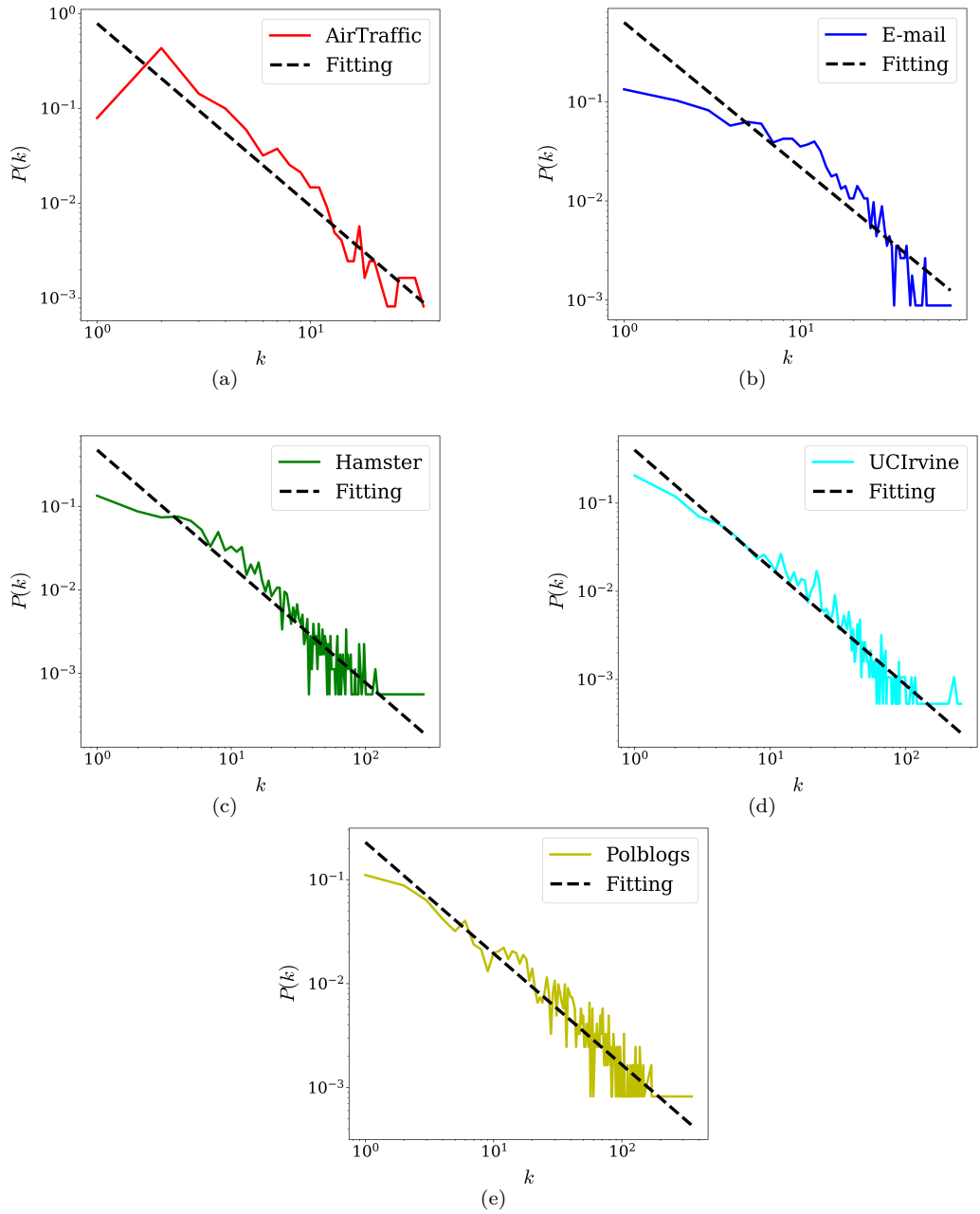

Figure 3: Fittings of the power-law exponents $\gamma$ estimated by the least-squares method for degree distributions $P(k) \sim k^{-\gamma}$ in real networks. (a) $\gamma = 1.92$ for AirTraffic. (b) $\gamma = 1.46$ for E-mail. (c) $\gamma = 1.39$ for Hamster. (d) $\gamma = 1.33$ for UCIrvine. (e) $\gamma = 1.07$ for Polblogs.

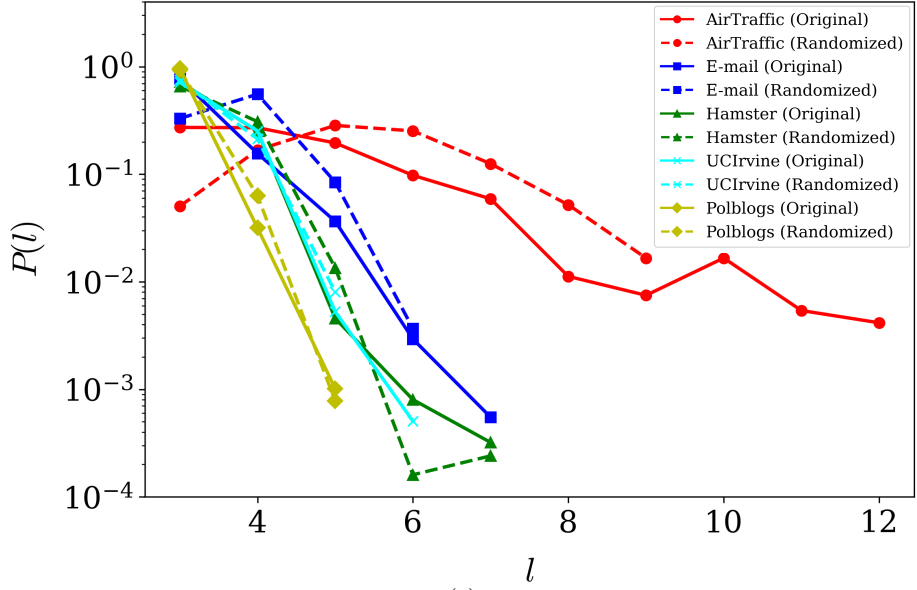

(a)

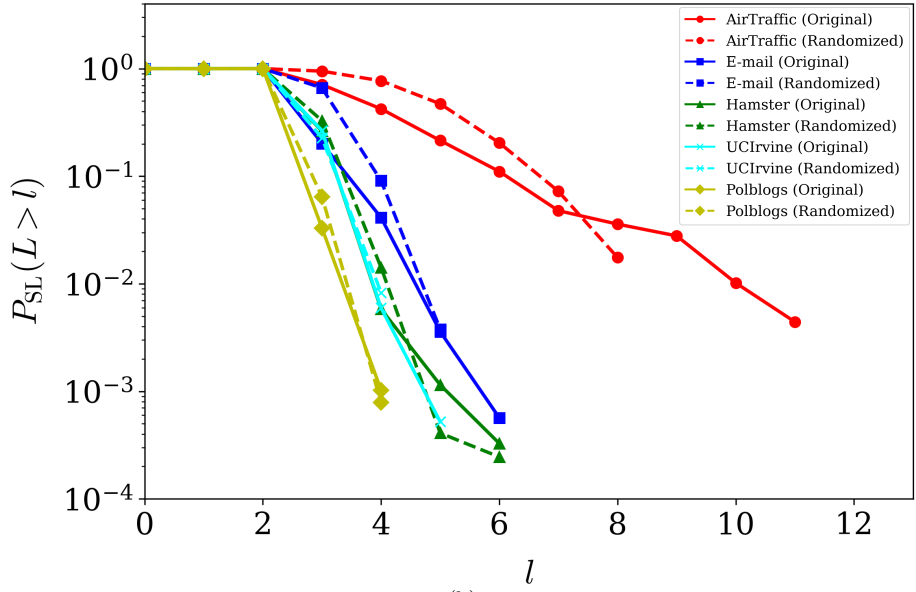

(b)

Figure 4: Distributions of the lengths $l$ of the shortest loops for real and the randomized networks. The results for AirTraffic, E-mail, Hamster, UCIrvine, and Polblogs are marked by red circles, blue squares, green triangles, cyan crosses, and gold diamonds, with solid lines. Dashed lines denote the results for the randomized networks. (a)(b) In comparing the tails of the distributions $P(l)$ and the cumulative distributions $P_{\text{SL}}(L > l)$, E-mail, Hamster, UCIrvine, and Polblogs contain shorter loops, except AirTraffic.

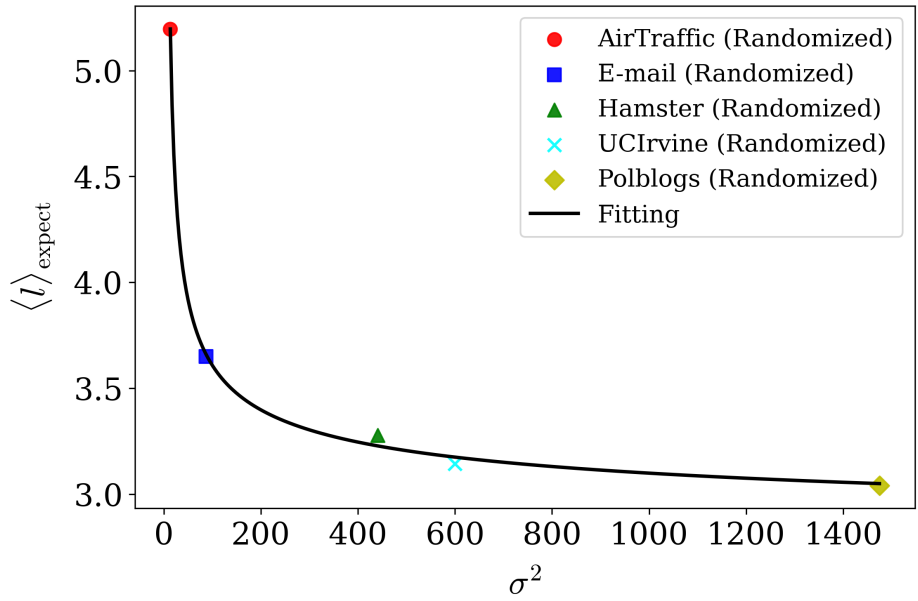

Figure 5: Dependence of the average length $\langle l \rangle_{\text{expect}}$ of the shortest loops in the randomized networks on the variance $\sigma^2$ of degree distributions $P(k)$ as the pure effect. The results for AirTraffic, E-mail, Hamster, UCIrvine, and Polblogs are marked by red circles, blue squares, green triangles, cyan crosses, and gold diamonds, respectively. As $\sigma^2$ increases, $\langle l \rangle_{\text{expect}}$ becomes smaller. The fitting denoted by black solid line is obtained by using the function $a_3/\log(a_1 x^2 + a_2) + a_4$, whose parameters $a_1, a_2, a_3$, and $a_4$ are estimated by the least-squares method.

*3.2. Dependence of the average length of the shortest loops on degree distributions*

In Subsection 3.1, we show that networks with the larger variance $\sigma^2$ of degree distributions $P(k)$ contain more triangles. Since the larger $\sigma^2$ corresponds to a wider $P(k)$ in approaching SF from regular networks, this suggests that the lengths of the shortest loops depends on the width of $P(k)$ as the pure effect with randomizations. Thus, we investigate the relation between the variance $\sigma^2$ of $P(k)$ and the average length $\langle l \rangle$ of the shortest loops by continuously changing degree distributions from power-law, nearly Poisson, to nearly unimodal through varying the value of parameter $\nu$.

We numerically calculated the distributions and the average length $\langle l \rangle$ of the shortest loops in networks with various $P(k)$ for $N = 1000, 5000, 10000$, and $50000$. Figure 6 shows the distributions of the lengths $l$ of the shortest loops for $N = 50000$. We obtain similar results for other sizes $N$. Figure 6(a) shows the distributions $P(l)$ for varying the value of $\nu$. The peak of $P(l)$ appears at left for $\nu = 1$ with gold line, and shifts to right as $\nu$ decreases for other light blue, orange, purple, green, blue, and red lines. Figure 6(b) similarly shows the cumulative distributions $P_{\mathrm{SL}}(L > l)$ shift to right as $\nu$ decreases. This means that regular networks ($\nu = -100$) contain longer shortest loops than SF networks ($\nu = 1$), and that ER random graphs ($\nu = -1$) are the intermediate between them. From those results, the shortest loops become longer as degree distributions are narrower in Figure 1. In Figure 6(b), the values of $P_{\mathrm{SL}}(L > l)$ obtained in our results (shown by circles) slightly deviate from the theoretical estimations by using the simpler approach (solid line) [23]. The differences between our results and the theoretical estimations [23] may be from that the recursive calculations based on generating functions assume locally tree-like structures, and hence are not valid for a network with short loops [4, 5]. Furthermore, since the approach in [23] is node-centric whereas ours is edge-centric, these methods differ for counting loops. In the node-centric case, a shortest loop can be counted multiple times per node, while in the edge-centric case, it can be counted multiple times per edge. However, the multiplicities are probably different between the two cases, which may also affect the slight discrepancy between our results and the theoretical estimations [23]. Instead of the simpler approach [23], the detailed approach [23] may give a better approximation, however, it practically intractable because of the huge combinational computations.

Figure 7 shows a monotone decreasing of $\langle l \rangle$ as $\sigma^2$ increases for $N =$

1000, 5000, 10000, and 50000. We also include the results for $\nu = 0.8, 0.9$ and 0.97 to interpolate between the points. It is common that the smaller variance $\sigma^2$ corresponds to larger $\langle l \rangle$. In other words, networks with narrower $P(k)$ contain longer shortest loops in order from SF networks, ER random graphs, to regular networks. We also fit them with the function $a_3/\log(a_1 x^2 + a_2) + a_4$ by the least-squares method, but further examination will be required for a more detailed understanding.

The narrowest degree distribution $P(k)$ corresponds to regular networks. As a special case of $d$-regular networks, Ramanujan graphs are known to be highly tolerant to bisections and to exhibit a large girth of order $O(\log_{d-1} N)$ in the limit $N \to \infty$ [26, 27]. In addition, random regular networks are known to asymptotically approach Ramanujan graphs as their size $N$ increases [28]. Figure 8 shows the average length $\langle l \rangle \approx \log_{d-1} N$, when $\nu \ll 0$. We give an attention to estimating $\langle l \rangle \approx \log_{d-1} N$ in random regular networks with unimodal degree distributions. On the other hand, in more general random networks with arbitrary $P(k)$, the average length $\langle l_{\text{sp}} \rangle$ of the shortest paths has already been analyzed [44] as follows:

$$\langle l_{\text{sp}} \rangle \simeq \frac{\ln N}{\ln \mu}, \tag{3}$$

$$\mu = \frac{\langle k^2 \rangle - \langle k \rangle}{\langle k \rangle}, \tag{4}$$

where $\mu$ is the excess mean degree. In the case of $\langle k \rangle = d$ for regular networks, we have $\mu = d - 1$. Then, we have $\langle l_{\text{sp}} \rangle \simeq \log_{d-1} N$, which coincides with our estimated $\langle l \rangle$ as the special case of unimodal degree distributions. This point is discussed in Conclusion. As shown in Figure 8, regular networks (red line at $\nu = -100$) contain longer shortest loops as $N$ increases, this exhibits the same scaling behavior as Ramanujan graphs. Thus, it is suggested that such loop lengths are not too large to maintaining the connectivity even against intentional attacks. We emphasize that the order of decreasing $\sigma^2$ of $P(k)$ is from SF networks, ER random graphs, to regular networks coincide with the order of both increasing $\langle l \rangle$ and improving the robustness of connectivity against attacks [12]. Therefore, the variance $\sigma^2$ of $P(k)$ is dominant on not only the robustness but also the average length $\langle l \rangle$ of the shortest loops in the wide class of random networks.

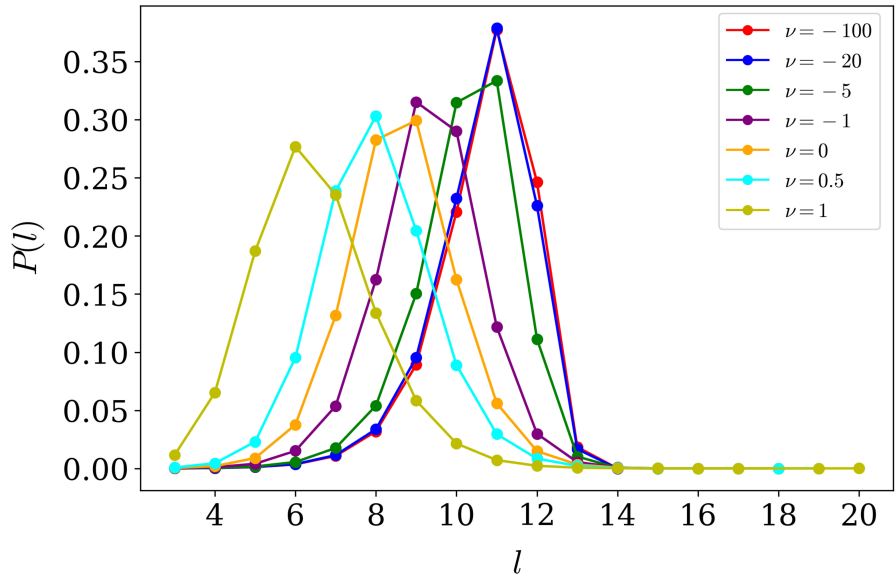

(a)

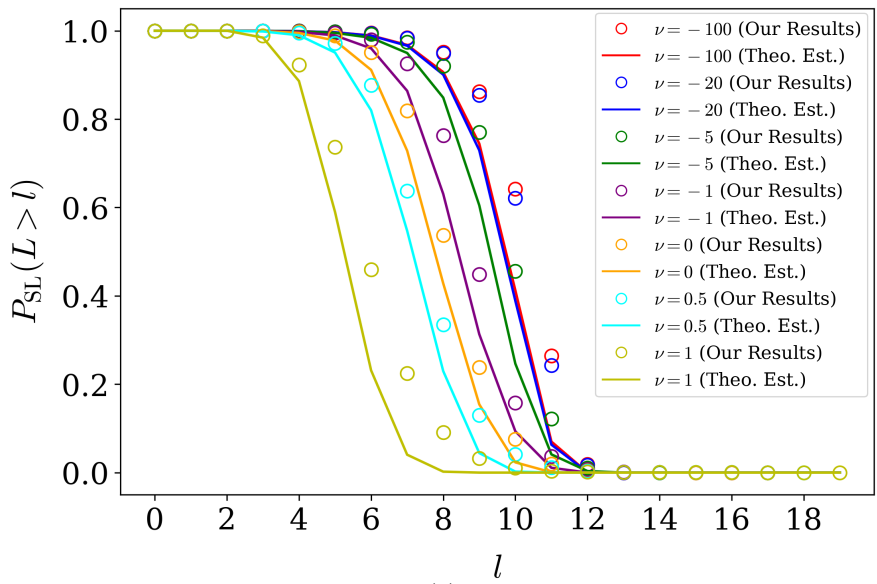

(b)

Figure 6: Distributions of the lengths $l$ of the shortest loops for the parameter $\nu$ in the attachment probability proportional to $k_i^\nu$. As $\nu$ decreases, networks contain longer shortest loops. (a) The peak of distribution $P(l)$ shifts to right as $\nu$ decreases. This indicates that regular networks (red line at $\nu = -100$) contain longer shortest loops than SF networks (gold line at $\nu = 1$), while ER random graphs (purple line at $\nu = -1$) are the intermediate between them. (b) Similarly, the cumulative distribution $P_{\text{SL}}(L > l)$ shifts to right as $\nu$ decreases. Our results (open circles) slightly deviate from the theoretical estimation (solid lines) by using the simpler approach [23].

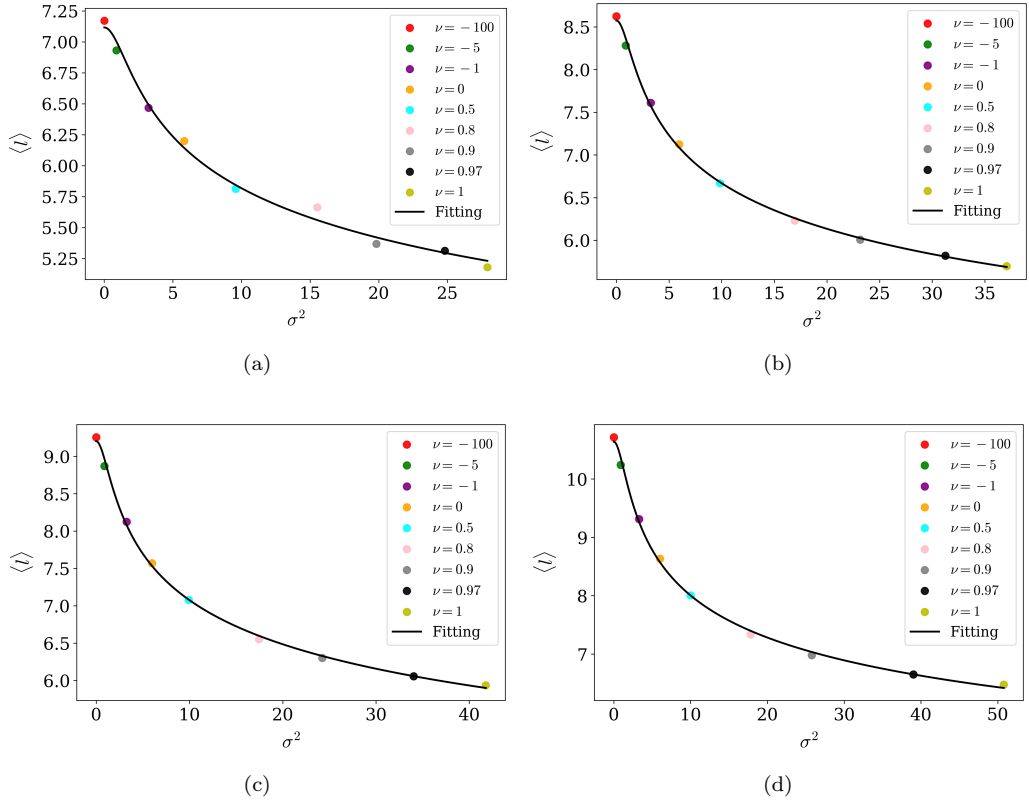

Figure 7: Dependence of the average length $\langle l \rangle$ of the shortest loops on the variance $\sigma^2$ of degree distributions $P(k)$ in networks with the size (a) $N = 1000$, (b) $N = 5000$, (c) $N = 10000$, and (d) $N = 50000$. Smaller values of $\sigma^2$ give larger $\langle l \rangle$, since narrower $P(k)$ contain longer loops especially in regular networks at $\nu = -100$ (red circle plots). Note that the cases of $\nu = -20$ are omitted because circle plots are almost overlapping with red ones at $\nu = -100$, which correspond to regular networks. The fittings denoted by black solid lines are obtained by using the function $a_3/\log(a_1 x^2 + a_2) + a_4$, whose parameters $a_1, a_2, a_3$, and $a_4$ are estimated by the least-squares method.

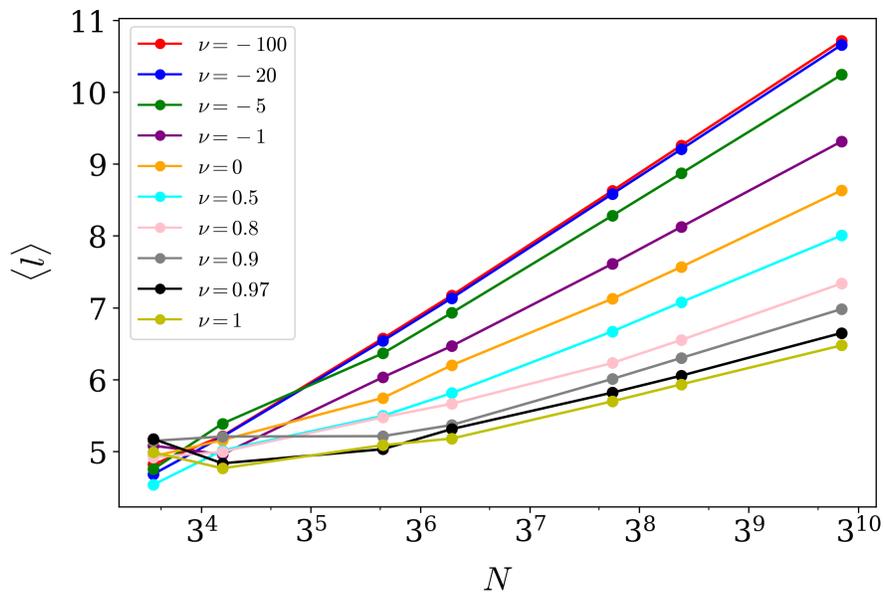

Figure 8: Dependence of the average length $\langle l \rangle$ of the shortest loops on the size $N$ under $\langle k \rangle \approx 4$. In particular, $\langle l \rangle$ becomes larger as $N$ increases, especially $\langle l \rangle \approx \log_3 N$ in $\nu \ll 0$ as shown by red straight lines at $\nu = -100$, which correspond to regular networks. The cases of $\nu = 1$ and $-1$ correspond to SF networks and ER random graphs, respectively.

## 4. Conclusion

We have investigated the lengths of the shortest loops in various networks with continuously changing degree distributions generated by using GN [36] and IPA models [37]. Our numerical analysis revealed that networks with narrower degree distributions contain longer length of the shortest loops in the wide class of random networks. This coincides with improvement of the robustness of connectivity against attacks as degree distributions become narrower [12]. Moreover, in random regular networks, the average length $O(\log_{d-1} N)$ of the shortest loops is shown as similar to Ramanujan graphs [26, 27]. Although a combination of large holes and stronger connectivity seems to be a contradiction, they suggest that the average length of the shortest loops of $O(\log_{d-1} N)$ is not too large for maintaining the connectivity even against intentional attacks. One of the reason may be caused by that a Ramanujan graphs has good property with large expander, which means the tendency not to disconnect into two parts.

On the other hand, further studies remain to clarify more detail relation between the variance $\sigma^2$ of degree distributions and the average length $\langle l \rangle$ of the shortest loops. We also fit them with the function $a_3/\log(a_1 x^2 + a_2) + a_4$ by the least-squares method, but further investigation is needed theoretically. The average length of the shortest paths is given by $\langle l_{\text{sp}} \rangle \simeq \ln N / \ln \mu$ in Eq (3) [44]. From Eq (4) at a fixed $\langle k \rangle$, $\mu$ is proportional to $\langle k^2 \rangle$ and to $\sigma^2 = \langle k^2 \rangle - \langle k \rangle^2$. Thus, $\langle l_{\text{sp}} \rangle$ decreases monotonically as $\sigma^2$ increases. Although the shortest paths and the shortest cycles are deeply related, they are not exactly equal. It remains unclear whether $\langle l_{\text{sp}} \rangle \neq \langle l \rangle$ or not. For this question, the multiplicities in the edge-centric or node-centric may be related as mentioned in Subsection 3.2. Apart from our attention to random regular networks with $\langle l \rangle \approx \log_{d-1} N$, more detailed investigations through numerical fitting to $\ln N / \ln \mu$ still remains as a future work. It may also be useful to estimate, e.g., the explicit function of $\langle l \rangle$ for the average degree $\langle k \rangle$ [45], and to investigate articulation points and bredges [46, 47, 48], as well as possible extensions of the framework to leader selection [49] in distributed processing.

## CRediT authorship contribution statement

**Kiri Kawato**: Methodology, Software, Investigation, Data curation, Writing - original draft. **Yukio Hayashi**: Conceptualization, Writing - review & editing, Supervision.

**Declaration of competing interest**

The authors declare that they have no known competing financial interests or personal relationships that could have appeared to influence the work reported in this paper.